# Interface Structure of Graphene on SiC( 000 $\bar{1}$ )

N. Srivastava, Guowei He, Luxmi, and R. M. Feenstra[*]
Dept. Physics, Carnegie Mellon University, Pittsburgh, PA 15213

**Abstract**
Graphene films prepared by heating the SiC( 000 $\bar{1}$ ) surface (the *C-face* of the {0001} surfaces) in vacuum or in a Si-rich environment are compared. It is found that different interface structures occur for the two situations. The former yields a well known 3×3 reconstructed interface, whereas the latter produces an interface with √43×√43-R±7.6° symmetry. This structure is shown to contain a graphene-like layer with properties similar to the 6√3×6√3-R30° "buffer layer" that forms on the Si(0001) surface (the *Si-face*).

Production of epitaxial graphene on SiC{0001} is being actively pursued by many research groups because of the potential application of graphene for novel electronic devices.[1] In terms of the graphene/SiC interface structure, the situation is quite well understood for the SiC(0001) surface (the Si-face): the interface consists of a C-rich layer having 6√3×6√3-R30° symmetry (denoted 6√3 for short) that is covalently bonded to the SiC.[2,3] This layer acts as an electronic "buffer" between the SiC and overlying graphene layers.[4] For the SiC( 000 $\bar{1}$ ) surface (the C-face) the situation is less well understood. Contradictory results have been obtained in different studies: most studies reveal a 3×3 and/or 2×2 interface structure with graphene layers being *weakly bound* to the underlying SiC, [3,5] whereas a single study [6] revealed an interface layer that was *strongly bound* to the SiC (i.e. similar to the Si-face situation). The sample used in the latter study differed from those of the former in terms of its preparation (furnace grown vs. vacuum prepared), so as suggested by Hass et al., it is possible that their interface structures could be different.[7]

We demonstrate in this work that, indeed, the interface between graphene and SiC( 000 $\bar{1}$ ) depends on the means of forming the graphene. For formation in vacuum, we observe a 3×3 interface structure, in agreement with that seen by many other groups.[1,3,5] But when the graphene is formed in a Si-rich environment, produced either by ≈$10^{-4}$ Torr of disilane or using 1 atm of purified neon, we observe a C-rich interface layer with structure √43×√43-R±7.6°. This interface structure is somewhat similar to the 6√3×6√3-R30° interface layer of the Si-face, but with the supercell for the C-face interface being rotated by only ±7.6°, rather than 30°, relative to the SiC axes. With this interface layer lying directly on the C-face SiC, it is found to form a buffer layer between the SiC and the graphene, similar to that modeled theoretically by Varchon et al.[4]

Experiments are performed on nominally on-axis, *n*-type 6H-SiC or semi-insulating 4H-SiC wafers purchased from Cree Corp, with no apparent differences between results for the two types of wafers. The wafers are cut into 1×1 cm$^2$ samples. To remove polishing damage, the samples are heated in either 1 atm of hydrogen at 1600°C for 3 min or 5×$10^{-5}$ Torr of disilane at 850°C for 5 min. In the same chamber, graphene is formed by heating either in vacuum ($10^{-8}$ Torr), in $10^{-6}$ – $10^{-4}$ Torr of disilane, or in 1 atm of neon.[8] In the latter case the neon is cryogenically purified by flowing liquid $N_2$ between the walls of the double-walled chamber, thus inhibiting unintentional oxidation of the C-face SiC which was a problem in prior work using argon.[9] Characterization by low-energy electron diffraction (LEED) is performed *in situ* in a connected ultra-high-vacuum chamber. After transferring the samples through air, further characterization is performed using an Elmitec III low-energy electron microscope (LEEM).

---

[*] feenstra@cmu.edu

Figure 1 illustrates, using LEED, the different surface (and interface) structures that we find on SiC($000\bar{1}$) surfaces prepared in either vacuum or disilane. Figure 1(a) shows the well-known[1,3,5] 3×3 pattern that forms on a vacuum-prepared surface. Faint graphene streaks are also observed in that pattern, corresponding to a thin graphene coverage [<0.5 monolayer (ML)], with these streaks arising from the rotational disorder that occurs on this C-face surface.[10,11] For samples heated for greater temperature or time, the 3×3 spots diminish and the graphene streaks grow in intensity. The 3×3 pattern itself is known to be a stable reconstruction of the ($000\bar{1}$) surface, and it has also been reported to persist at the graphene/SiC interface.[5]

For graphene formed on the C-face at disilane pressures near $10^{-6}$ Torr, we find the same LEED pattern as for vacuum (and also similar LEEM reflectivity curves).[12] However, at disilane pressures near $10^{-4}$ Torr we find qualitative differences between vacuum- and disilane-prepared surfaces. Figure 1(b) shows a LEED pattern from a surface prepared by heating in 5×$10^{-5}$ Torr of disilane at 1270°C. Weak graphene streaks are visible along with the primary SiC spots, as marked, and a complex arrangement of additional spots is apparent. Analysis of these spots is shown in Fig. 1(c).[13] The pattern can be perfectly indexed using a supercell on the SiC with edges extending along (6,1) and (-1,7) of the SiC 1×1 cells. In conventional notation this structure would be expressed as a matrix with columns (6,1) and (-1,7), and in a more compact notation we denote this structure as √43×√43-R±7.6° (or √43 for short) with the 7.6° = $\tan^{-1}(\sqrt{3}/13)$ being the rotation of the supercell relative to the SiC. Approximately 8×8 unit cells of graphene fit within this super cell (with 2.4% mismatch, using room temperature lattice constants $a_{SiC}$=0.3080 nm and $a_{graphite}$=0.2464 nm).

The results of Fig. 1 demonstrate that different surface structures exist for the disilane-prepared SiC($000\bar{1}$) compared to a vacuum-prepared surface. Additional information comes from LEEM studies. However, for those studies it is necessary for us to transfer the samples through air to the LEEM instrument. This air exposure causes the √43 pattern to disappear, being replaced by a √3×√3-R30° pattern as shown in Fig. 1(d). Intensity vs. energy characteristics of the √3 spots, e.g. Figs. 1(e) and 1(f), agree semi-quantitatively with those for the well-known $Si_2O_3$ silicate structure on the ($000\bar{1}$) surface.[14] However, the selected-area LEED results presented below demonstrate that a graphene or graphene-like layer terminates the surface, so we conclude that that layer has the silicate (or some other oxidized SiC) structure *below* it. This modification of the ($000\bar{1}$) surface structure also affects its LEEM reflectivity curves, as described below.

As an introduction to our LEEM results, we first present data in Fig. 2 for vacuum-prepared C-face graphene.[9] Figure 2(a) shows a LEEM image displaying varying contrast due to different numbers of graphene layers on the surface. Measurements of the reflected intensity of the electrons as a function of their energy from specific surface locations B – E are shown in Fig. 2(b). These reflectivity curves display simple oscillatory behavior over the range 2 – 6 eV. As demonstrated by Hibino et al., the number of minima in the reflectance curves corresponds to the number of graphene layers on the surface.[15,16] On surfaces where the average thickness of the graphene film is low, some portions of the surface display reflectivity curves that are relatively featureless, as in curve A of Fig. 2(b). We interpret those areas as being devoid of ordered graphene (although a small amount of disordered, nano-crystalline graphite is present on those regions because of annealing of the sample as part of the LEEM preparation procedure[12]).

LEEM data for the same disilane-prepared sample described in Figs. 1(b) – 1(f) are shown in Fig. 3. The LEEM image of Fig. 3(a) displays areas with mainly two different contrasts, bright and dark, along with a mottled area with intermediate contrast. Similar such areas were found to extend over the entire sample. Curves B and C acquired from the bright area and mottled area, respectively, display a broad maximum in the reflectivity over 2 – 6 eV along with a distinct minimum at 6.4 eV. These curves have no analog in the vacuum-prepared sample, consistent with the presence of a new type of surface structure. Curve D displays a minimum near 3.5 eV, suggestive of a single graphene layer, and it also has a small minimum near 6.7 eV. Curve E displays only a single minimum near 2.8 eV (results like curve E only occur over ≈2% of the surface area).[17] Notably, selected-area diffraction acquired from both bright (like B) or dark (like D) areas both display diffraction spots with wavevector magnitude equal to that of graphene, as shown in Figs. 3(c) and (d). Thus, a graphene (or graphene-like) layer apparently extends over the entire surface. On other samples prepared in disilane with similar conditions as that of Fig. 3, the same set of reflectivity curves as seen in curves B – E of Fig. 3(b) are found, and we also find a reflectivity curve that is featureless over 2 – 6 eV as shown by curve A of Fig. 3(b). We associate this curve with the absence of graphene on the surface [i.e. similar to that seen by curve A of Fig. 2(b)].

Additional information on the identity of the various structures seen in Fig. 3 comes from studies of surface oxidation and annealing in the LEEM. As already mentioned above, the disilane-prepared surfaces are found to oxidize upon air exposure. However, we find that this oxidation of the surface proceeds more readily in some surface areas than others, so that further oxidation can be accomplished in the LEEM, as shown in Fig. 4. The surface there is one that was prepared by heating in a 1-atm purified neon atmosphere, and it displayed a complex *in situ* LEED pattern identical to Fig. 1(b). After transfer of the sample through air into the LEEM, data shown in Figs. 4(a) and 4(c) were obtained. Curves E and F of Fig. 4(c) in particular reveal single-ML-thick graphene. The surface was then exposed to $1\times10^7$ L of oxygen with the sample at ≈200°C after which it was briefly heated to 1000°C. The surface areas from which the reflectivity curves E and F were acquired were modified by this procedure, producing reflectivity curves as shown in Fig. 4(d). Those curves show minima near 7.1 eV, similar to curve D of Fig. 3(b). We thus find that this sort of characteristic is indeed associated with single ML graphene, although upon oxidation its structure changes. Selected area diffraction from various locations on the surface of Fig. 4 (not shown) are found to reveal spots with wavevector equal to that of graphene.

The change in the appearance of reflectivity curves E and F of Fig. 4 due to oxidation is very similar to what has been previously reported for the SiC(0001) surface, associated with a *decoupling* of the 6√3 buffer layer from the underlying SiC due to oxidation or hydrogenation.[18,19,20] The decoupled buffer layer then assumes a diffraction pattern identical to that of graphene, i.e. it becomes a graphene layer. We interpret our results for the ($000\bar{1}$) surface in exactly the same way: The unique reflectivity curves B and C of Fig. 3(b) are attributed to a buffer-like layer on the surface, and we associate the √43 diffraction pattern to this same layer. We have not been able to observe that pattern in the LEEM, since that part of the surface becomes oxidized during transfer to the LEEM. However, we do find a clear correlation between the intensity of the √43 LEED pattern from various samples and the prevalence of the surface areas with reflectivity curves like B and C of Fig. 3(b) or A of Figs. 4(c) and 4(d), so we can confidently associate them.

With this association of the √43 pattern and the unique reflectivity curves, we argue that our data provides a compelling case for this structure being a graphene buffer layer on the ($000\bar{1}$) surface, analogous to the 6√3 layer on the (0001) surface: (i) we find that this layer, after oxidation, has a diffraction pattern identical to that of graphene; (ii) reflectivity curves for graphene on the √43 layer are modified by oxidation (decoupling), similar to what occurs on the (0001) surface;[18] (iii) the reflectivity curves for the oxidized √43 layer itself bear some resemblance to those for a decoupled 6√3 layer (e.g. Fig. 3(c) of Ref. [21]); (iv) the LEED pattern of the unoxidized √43 layer is also somewhat analogous to that of the 6√3 layer in that they both reveal a large-area superstructure on the SiC surface.

We emphasize that our identification of this new interface structure (graphene buffer layer) on the ($000\bar{1}$) surface does *not* rely on the assumption of one reflectivity minima in the reflectivity curve per monolayer of graphene (i.e. as in the early work of Hibino et al.)[15,16] since, indeed, recent results demonstrate that the situation is more complicated than that.[21] Rather, the reflectivity characteristics are just used as a means of distinguishing unique structures on the surface, and the relative graphene coverage for each structure has been assigned on the basis of which structure predominates on samples with given average thickness of graphene, i.e. curve B of Fig. 3(b) is seen on samples with thinnest graphene, followed by curve D for thicker graphene films, etc. (curve A is seen on surface areas devoid of graphene).

Regarding the *reason* for the different interface structures in a Si-rich environment compared to vacuum, the graphene formation conditions in the former case are expected to be closer to equilibrium, as argued by Tromp and Hannon,[22] so that kinetic limitations may lead to the absence of the √43×√43-R±7.6° structure in vacuum-prepared C-face graphene. Complete geometrical determination of the √43 structure (as well as the C-face 3×3 structure) is needed before a full understanding of its formation can be achieved. In any case, the fact that we obtain the same √43 interface structure using either disilane or a purified neon environment (and it was also seen previously in small areas on an argon-prepared surface)[9] demonstrates that the hydrogen from the disilane is unimportant. Separately, it is interesting to note that in Figs. 3(c) and 3(d) most of the graphene diffraction spots (both the buffer layer and the overlying graphene) are rotationally misaligned from the SiC spots by angles in the range ±10°, the same range as found in studies of thick graphene on C-face SiC.[9,10,11] Thus, the √43 buffer structure observed here may play a role in the origin of this commonly observed rotational misalignment.[23]

We gratefully acknowledge support from the National Science Foundation (grant DMR-0856240).

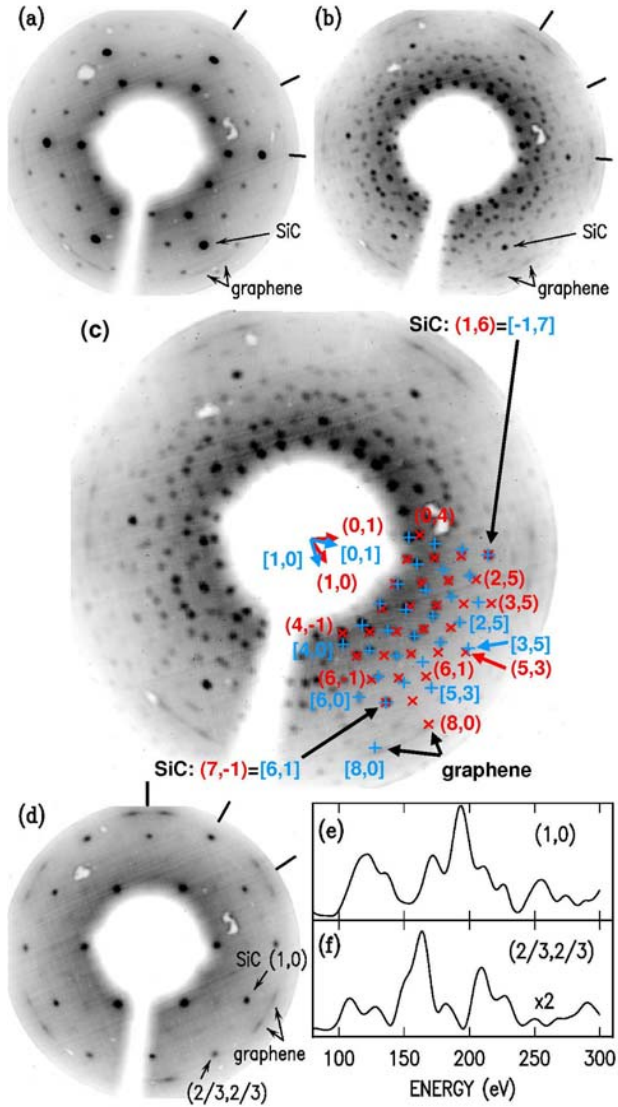

FIG. 1. LEED results obtained at 100 eV from 6H-SiC($000\bar{1}$) surfaces prepared under various conditions: (a) heated in vacuum at 1130°C for 15 min yielding an average graphene thickness of <0.5 ML, (b) heated in $5\times10^{-5}$ Torr of disilane at 1270°C for 15 min yielding an average graphene thickness of 0.6 ML (not counting the $\sqrt{43}$ buffer layer), (c) same as (b) but with LEED spots marked (see text), (d) same sample as (c) but after air exposure, (e) and (f) intensity vs. energy of the spots indicated in (d). The thin black lines in (a), (b) and (d) indicate two 30° ranges of angles, for reference.

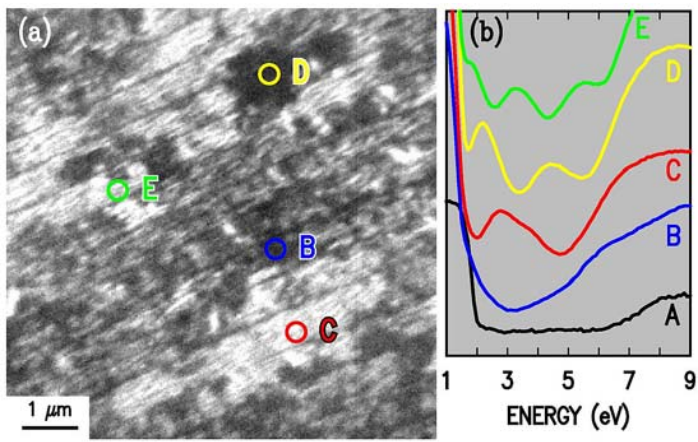

FIG. 2. Results for graphene on 6H-SiC($000\bar{1}$) prepared by heating in vacuum at 1100°C for 20 min yielding an average graphene thickness of 2.0 ML of graphene. (a) LEEM image at an electron beam energy of 3.3 eV. (b) Curves B-E show the intensity of the reflected electrons acquired from the circular areas marked in (a), and curve A shows data from a different sample with less graphene coverage prepared by heating at 1170°C for 15 min [similar to that of Fig. 1(a)].

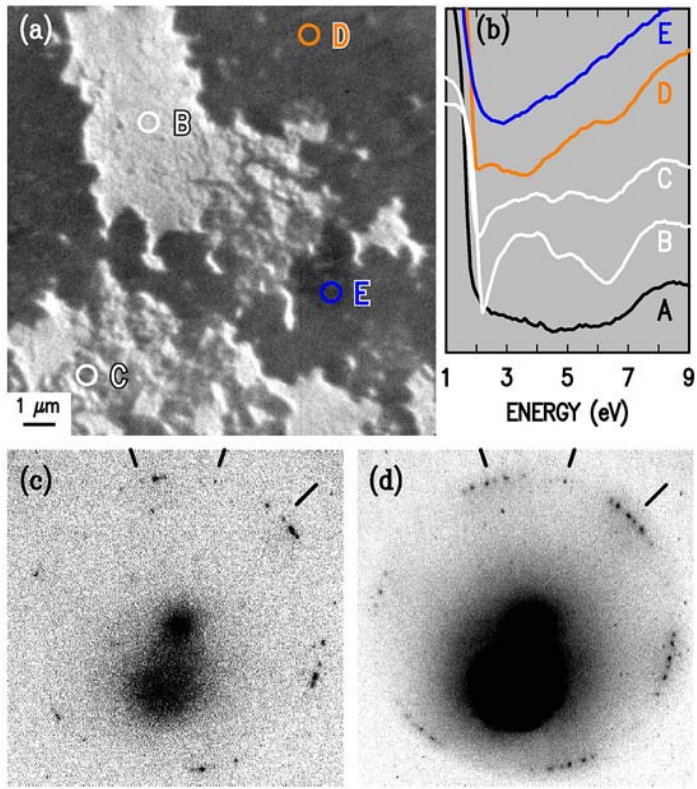

FIG 3. Results for graphene on 6H-SiC($000\bar{1}$) prepared by heating in $5\times10^{-5}$ Torr of disilane at 1270°C for 15 min [same sample as Fig. 1(b) and 1(c)], after air exposure. (a) LEEM image acquired at electron beam energy of 4.5 eV. (b) Curves B – E show the intensity of the reflected electrons acquired from the circular areas marked in (a), and curve A shows data from a different sample with less graphene coverage prepared by heating in $6\times10^{-5}$ Torr at 1310°C for 10 min. (c) and (d) Selected-area patterns acquired with a 2-μm aperture at 44 eV in the LEEM, from locations indicated by B and D, respectively, in panel (a). The thin black lines indicate two 30° ranges of angles, for reference.

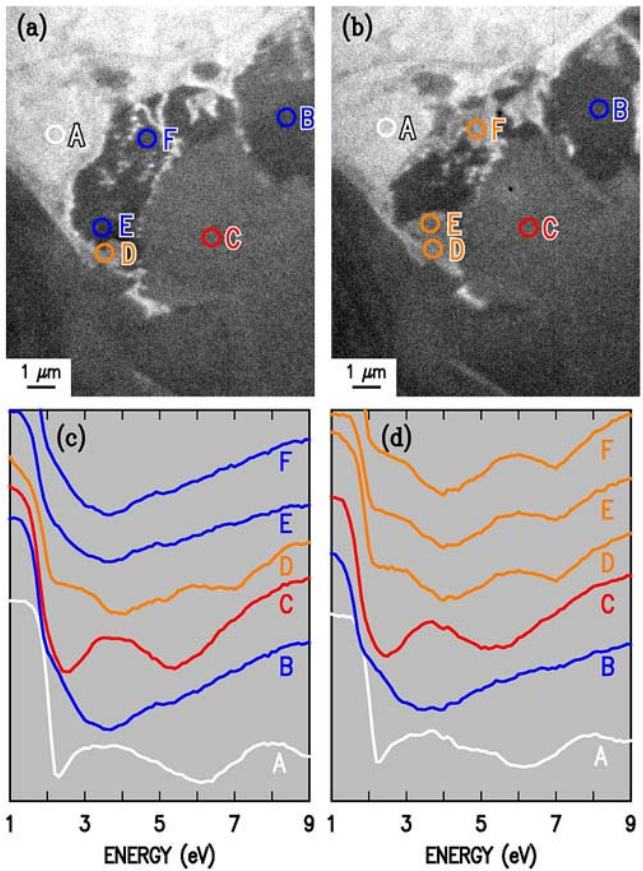

FIG 4. Results for graphene on 4H-SiC($000\bar{1}$) prepared by heating in 1 atm of neon at 1450°C for 10 min. (a) and (b) LEEM images at 3.1 eV, before and after oxidation of the sample, respectively. (c) and (d) Reflectivity curves acquired from the circular areas marked in (a) and (b), respectively.

[23] It should be noted, however, that the ±10° spread is mainly observed for vacuum-prepared SiC, for which the interface (at least for very thin films) is believed to be the 3×3 and/or 2×2 structures. But, the buffer layer could still be present at the elevated temperatures used to form the multi-layer films.